\documentclass[aps,pre,floats,reprint,onecolumn,showpacs,superscriptaddress]{revtex4}

\usepackage{graphicx,epsfig}
\usepackage{times}
\usepackage{graphics,dcolumn,bm,fleqn,float}
\usepackage{amssymb,amsmath,multirow,rotate,color}
\begin{document}

\title{Spreading of Persistent Infections in Heterogeneous Populations}

\author{J. Sanz}

\affiliation{Institute for Biocomputation and Physics of Complex
Systems (BIFI), University of Zaragoza, Zaragoza 50009, Spain}

\author{L. M. Flor\'{\i}a}

\affiliation{Institute for Biocomputation and Physics of Complex
Systems (BIFI), University of Zaragoza, Zaragoza 50009, Spain}

\affiliation{Departamento de F\'{\i}sica de la Materia Condensada,
Universidad de Zaragoza, Zaragoza E-50009, Spain}

\author{Y. Moreno}

\email{yamir.moreno@gmail.com}

\affiliation{Institute for Biocomputation and Physics of Complex
Systems (BIFI), University of Zaragoza, Zaragoza 50009, Spain}

\affiliation{Departamento de F\'{\i}sica Te\'orica. Universidad de
Zaragoza, Zaragoza E-50009, Spain}

\date{\today}

\begin{abstract}

Up to now, the effects of having heterogeneous networks of contacts have been studied mostly for diseases which are not persistent in time, i.e., for diseases where the infectious period can be considered very small compared to the lifetime of an individual. Moreover, all these previous results have been obtained for closed populations, where the number of individuals does not change during the whole duration of the epidemics. Here, we go one step further and analyze, both analytically and numerically, a radically different kind of diseases: those that are persistent and can last for an individual's lifetime. To be more specific, we particularize to the case of Tuberculosis' (TB) infection dynamics, where the infection remains latent for a period of time before showing up and spreading to other individuals. We introduce an epidemiological model for TB-like persistent infections taking into account the heterogeneity inherent to the population structure. This sort of dynamics introduces new analytical and numerical challenges that we are able to sort out. Our results show that also for persistent diseases the epidemic threshold depends on the ratio of the first two moments of the degree distribution so that it goes to zero in a class of scale-free networks when the system approaches the thermodynamic limit.

\end{abstract}


\maketitle

\section{Introduction}

Disease spreading has been the subject of intense research
since long time ago \cite{MayBook,Cambridge,Murray}.  Our current knowledge comprises mathematical models that have allowed to better
understand how an epidemic spreads and to design more efficient immunization and vaccination policies \cite{MayBook,Cambridge,
Murray}. These models have gained in complexity in recent years capitalizing on data collections which have provided
information on the local and global patterns of relationships in the population \cite{geisel,guimera,colizza}. In
particular, with the advent of modern computational resources and tracking systems, it is now feasible to 
contact-trace the way the epidemic spreads or at least to predict the paths that a given pathogen might follow.  
In this way, some of the assumptions at the basis of the theoretical models that were difficult to
test -the backbone through which the diseases are transmitted- are now more accurately incorporated into epidemiological models \cite{egamstw04,cpv07,glmp08,cv08,mam09}.

Strikingly, the systems on top of which diseases spread show common nontrivial topological and statistical properties \cite{siam,PhysRep}.  A large number of networks of contacts in real-world social, biological and  technological systems have been found to be best described by the so-called scale-free (SF)
networks. In SF networks, the number of contacts or connections of a node with
other nodes in the system, the degree (or connectivity) $k$, follows a power
law distribution, $P(k) \sim k^{-\gamma}$.  Recent studies have shown that the SF topology has a great impact on the dynamics and function of the system under
study \cite{cnsw00,ceah01,siam,PhysRep}. The reason is that, at variance with homogeneous or regular networks, SF architectures are a limiting case of heterogeneity where the connectivity
fluctuations diverges if $2<\gamma\leq 3$ as the system size tends to infinity (the thermodynamic limit). This means that there are nodes in the network which has an eventually unbounded number of connections compared to the average degree. Examples of such networks include the Internet \cite{falou99,calda00}, the
world-wide-web (WWW) \cite{www99}, food-webs, and metabolic or protein networks \cite{strog01,PhysRep}.  

In the context of disease spreading, SF networks lead to a vanishing 
epidemic threshold in the limit of infinite population when $\gamma \leq 3$
\cite{Vespignani,MaySci,mpv02,newman02}.  This is because the ratio
$\langle k\rangle / \langle k^2\rangle$ determines the epidemic threshold
above which the outbreak occurs.  When $2 < \gamma \leq 3$, $\langle k\rangle$
is finite while $\langle k^2\rangle$ goes to infinity, that is, the
transmission probability required for the infection to spread goes to zero. Moreover, the previous result holds both for the Susceptible-Infected-Susceptible (SIS) and Susceptible-Infected-Removed (SIR) epidemiological models \cite{Vespignani,MaySci,mpv02}. 

In this paper, we will deal with a different kind of diseases -those that are persistent in time and shows a latent period that can be as large as an individual's lifetime. Our first aim is to enlarge the epidemiological framework for complex networks reported previously for the SIS model \cite{Vespignani} and proposed as well for the SIR model \cite{MaySci,mpv02} by integrating the spreading dynamics of persistent diseases within it. With this purpose, we consider a variation of the Susceptible-Exposed-Infected-Removed model \cite{MB} on complex  heterogeneous networks. As we will see, this kind of dynamics introduces new challenges essentially different to other successfully treated before. In particular, as the latent period is high enough, we shall work with an open system in which new individuals are born and others dead for causes not directly related to the spreading of the disease. We present novel analytical and numerical methods that allow us to obtain the epidemic threshold for this kind of disease in heterogeneous populations. Moreover, we introduce a numerical method that is well-suited to deal with the kind of problems we face. Our results point out that also for persistent infections the virtually unbounded fluctuations of the degree distribution have an important enhancing effect on the epidemic spreading.

\section{The model}

To be more precise and without loss of generality, we will particularize our model to one case of persistent infection -the most threatening one which is tuberculosis (TB). TB is an old disease whose world-wide prevalence had been diminishing even before vaccination and prophylaxis strategies were firstly accomplished \cite{Wilson,Styblo,Daniel}. Its recent return in developing countries, mainly in Southeast Asia, have attracted renewed interest in it. The current world estimate of prevalence is about 33\% while the number of deaths per year that it is causing reaches more than 3 million people  \cite{Bleed}. Depending of the kind and the intensity of immune response that the host immune system performs after initial infection with M. tuberculosis bacillus, the individual can suffer latent infection, (in which the bacteria are under a growth-arrest regime and the individual neither suffer any clinical symptom nor becomes infectious) or active infection, where the host suffers clinical symptoms and can transmit the pathogen by air \cite{Comstock,Styblodos}. Latently infected individuals can, generally after an immune-depression episode, reach the active phase. Estimating the probability of developing direct active infection after a contact, or alternatively, the lifetime's risk for a latent infected individual to evolve into the active phase, are not easy tasks. However, it is generally accepted that only 5-10\% of the infections directly produce active TB \cite{Comstock,Styblodos}, while the ranges concerning the estimation of typical ``half-life'' of latent state rounds about 500 years \cite{MB}.

The spreading dynamics of TB like diseases has been studied in recent years. However, to the best of our knowledge, these works assume the homogeneous mixing hypothesis, that is, a perfectly homogeneous system in which all individuals are dynamically equivalent. As mentioned above, many of the systems on top of which diseases spread, are better described by scale-free connectivity patterns. Therefore, in what follows, our main objective will be to assemble a basic model fitted for tuberculosis spreading that firstly takes into account the heterogeneity in the distribution of the networks of contacts. We note that the increasingly alarming situation about TB epidemiology evidences the need to increase the effort in TB research in a global way. In the context of the study of its epidemiology, new models must be developed in order to gain predictive skills, incorporating the recent theoretical advances referring to disease spreading on complex heterogeneous substrates as well as meta-population approaches and new computational tools for numerical analysis and simulation. In this sense, ours is a first contribution that addresses one of the most important parameters in epidemiological description: the epidemic threshold. 

Let us then introduce our model. We consider that individuals in the population are compartmentalized into three groups: healthy -$U(t)$-, infected but not infectious -\textit{or latently infected} $L(t)$- and sick individuals $T(t)$ which are infected and are infectious as well. The transition between these subpopulations proceeds in such a way that a healthy individual acquires the bacteria through a contact with an infectious subject with probability $\lambda$. In its turn, this newly infected individual may develop the disease directly with probability $p$. However, the most common case is the establishment of a dynamic equilibrium between the bacillus and the host's immune system, which allows the survival of the former inside the latter. When this happens, newly infected individuals become latently infected, because despite harboring the bacteria in blood, neither becomes sick nor is able to infect others.

On the other hand, after a certain period of time (which may be several years) and usually following an episode of immunosuppression, the balance between the bacterium and its host can be broken. In this case, the bacteria grow and the individual falls ill beginning to develop the clinical symptoms of the disease. In addition, if the infection attacks the lungs (pulmonary TB), the bacillus is present in the sputum, making the guest infectious.

The dynamics of the disease, in a well mixed population, is then described by the following system of nonlinear differential equations:
\begin{equation}
 \frac{dU(t)}{dt}=bN(t)-\lambda \beta U(t)\frac{T(t)}{N(t)}-\mu U(t),\nonumber
 \end{equation}
\begin{equation}
 \frac{dL(t)}{dt}=(1-p)\lambda \beta  U(t)\frac{T(t)}{N(t)}-(\mu+r) L(t),
  \label{a}
\end{equation}
\begin{equation}
 \frac{dT(t)}{dt}=p\lambda \beta  U(t)\frac{T(t)}{N(t)}+rL(t)-(\mu+\mu_{tb}) T(t),\nonumber
\end{equation}
in which:

\begin{description}
\item[$N(t)=U(t)+L(t)+T(t)$] represents the total population at time $t$,
\item [$\beta$] is the number of contacts per time unit,
\item [$\lambda$] is the probability that the bacteria is transmitted to a new host after a contact with an infectious subject,
\item [$b$] is the birth rate per capita and per unit time,
\item [$\mu$] is the natural death rate per capita and unit time,
\item [$\mu_{tb}$] is the rate of disease-related deaths per capita and unit time,
\item [$r$] is the transition frequency of latent infection (i.e., the probability that a latently infected individual becomes infectious),
\end{description} 

with the closure relationship:
\begin{equation}
 \frac{dN(t)}{dt}=(b-\mu) N(t) -\mu_{tb}T(t).
\end{equation}
The model above is a variation of the archetypal SIR model to which a fourth class has been added (latency class L). This kind of model has been largely treated in the literature in its well-mixed version, and it is frequently referred as SEIR model.  As a first step, in this work, we will identify the removed individuals mostly with dead ones, and therefore we do not consider the possibility of natural or medical recovery (this simplification is in part justified by the large latency period of infected individuals and the constant flow of newborns into the system). A more refined model would consist of introducing such eventual recovery fluxes in the model, as well as the possibility of further relapses (the so called endogenous reactivation). These phenomenologies might be important mainly for diseases (like TB) for which the only feasible treatment in many areas consists of supplying large series of antibiotics. Thinking on the tuberculosis case, further refinements, like the inclusion of varieties of less infectious extra-pulmonary diseases \cite{natmed95}, could also have consequences on the disease's dynamics.

\section{Structured Populations}
\subsection{Dynamics}

The previous system of differential equations describes the dynamics of the epidemics in the well-mixed case. However, as argued above, the number of contacts of a given individual in a population can vary, which is reflected in an heterogeneous distribution of the number of contacts in the system. To account for this fact, we next consider a structured population described by a connectivity distribution $P(k)$. The system of Eqs\ (\ref{a}) has to be modified accordingly. Assuming that all individuals with the same number of contacts, i.e., belonging to the same connectivity class $k$, are dynamically equivalent, the new system of differential equations are formulated for each degree class. Therefore, for a structured population, we have that: 
\begin{equation}
  N_k(t)=P(k)N(t),
\end{equation}
with:
\begin{equation}
 U_k(t)+L_k(t)+T_k(t)=N_k(t).
\end{equation}
Moreover, it is convenient to express the previous equations in terms of densities, also defined within each connectivity class:
\begin{equation}
 u_k(t)=\frac{U_k(t)}{N_k(t)},\nonumber
\end{equation}
\begin{equation}
l_k(t)=\frac{L_k(t)}{N_k(t)},
\end{equation}
\begin{equation}
 t_k(t)=\frac{T_k(t)}{N_k(t)},\nonumber
\end{equation}
so that the following closure relation for any value of $k$ is verified:
\begin{equation}
 u_k(t)+l_k(t)+t_k(t)=1\quad\quad\forall (k,t).
\end{equation}
On the other hand, the probability $\Theta$ that any given link points to an infectious individual is given by:
%
\begin{widetext}
\begin{equation}
 \Theta(t)=\frac{\sum_kkT_k(t)}{\sum_kkN_k(t)}=\frac{\sum_kkP(k)t_k(t)}{\langle k \rangle},
\end{equation} 
\end{widetext}
which leads to the following set of equations that describes the dynamics within each connectivity class:
\begin{equation}
 \frac{dU_k(t)}{dt}=bP(k)N-\lambda k\Theta(t) U_k(t)-\mu U_k(t),\nonumber
\end{equation}
\begin{equation}
 \frac{dL_k(t)}{dt}=(1-p) \lambda k\Theta(t) U_k(t)-(\mu+r) L_k(t),
 \label{h}
\end{equation}
\begin{equation}
 \frac{dT_k(t)}{dt}=p \lambda k\Theta(t)U_k(t)+rL_k(t)-(\mu+\mu_{tb}) T_k(t).\nonumber
\end{equation}
Finally, the number of individuals with connectivity $k$ evolves according to:
\begin{equation}
 \frac{dN_k(t)}{dt}=(b-\mu) N_k(t) -\mu_{tb}T_k(t)=(b-\mu-\mu_{tb}t_k)N_k(t).
 \label{d}
\end{equation}
At this point, and building on the previous equation, it is important to point out a feature of the model: the influence of the infection dynamics on the connectivity distribution $P(k)$. First, if we add the above equation for all $k$, we obtain that the total population evolves as:
\begin{equation}
 \frac{dN(t)}{dt}=(b-\mu) N(t) -\mu_{tb}\sum_kT_k(t)=\left(b-\mu-\mu_{tb}\sum_k{P(k)t_k}\right)N(t).
 \label{e}
\end{equation} 
However, if we substitute $N_k(t)=P(k)N(t)$ directly into Eq.\ (\ref{d}) and assume $P(k)$ to be constant, we would arrive to:

\begin{equation}
 P(k)\frac{dN(t)}{dt}=P(k)\left(b-\mu-\mu_{tb}t_k\right)N(t).\nonumber
\end{equation} 

The last expression is only compatible with Eq.\ (\ref{e}) under the unrealistic assumption that all connectivity classes have the same proportion of sick individuals. We must therefore assume that the distribution of connectivity is also a function of time: $P(k,t)$, and therefore:
\begin{equation}
 \frac{dN_k(t)}{dt}=\frac{d[P(k,t)N(t)]}{dt}=N(t)\frac{dP(k,t)}{dt}+P(k,t)\frac{dN(t)}{dt},
 \label{f}
\end{equation} 
so, if we substitute Eq.\ (\ref{f}) into Eq.\ (\ref{d})  we get:
\begin{equation}
 N(t)\frac{dP(k,t)}{dt}+P(k,t)\frac{dN(t)}{dt}=P(k)\left(b-\mu-\mu_{tb}t_k\right)N(t),\nonumber
\end{equation} 
expression from which, if we replace $dN(t)/dt$ from Eq.\ (\ref{e}), we get the temporal evolution of $P(k, t)$ as:
\begin{equation}
 \frac{dP(k,t)}{dt}=-P(k,t)\mu_{tb}\left[t_k(t)-\langle t_k\rangle (t)\right],
\end{equation} 
where
\begin{equation}
\langle t_k\rangle (t)=\sum_k{P(k,t)t_k(t)}.
\end{equation} 
Reformulating the equations in terms of densities using the definitions of the densities given above, Eqs.\ (\ref{h}) become
\begin{equation}
 \frac{du_k(t)}{dt}=b-u_k(t)(b+\lambda k\Theta(t) -\mu_{tb}t_k(t)),\nonumber
 \end{equation}
\begin{equation}
 \frac{dl_k(t)}{dt}=(1-p) \lambda k\Theta(t) u_k(t)-(b+r) l_k(t)+\mu_{tb}l_k(t)t_k(t),
 \label{g}
\end{equation}
\begin{equation}
 \frac{dt_k(t)}{dt}=p \lambda k\Theta(t)u_k(t)+rl_k(t)-(b+\mu_{tb})t_k(t)+\mu_{tb}t_k(t)^2.\nonumber
\end{equation}

\subsection{Evolution of the degree distribution}

At this point it is appropriate to point out one aspect that will hinder any numerical characterization of the epidemic threshold. Our main goal will be to calculate both numerically and analytically the critical value $\lambda_c$ beyond which the population presents an endemic proportion of sicks individuals. However, we expect $\lambda_c$ to be dependent on the ratio $\frac{\langle k \rangle}{\langle k^2 \rangle}$, which is in turn a function of the connectivity distribution $P(k)$. The degree distribution, as previously shown, changes in time as the dynamics of infection progresses. As we should see, we can handle this time dependence analytically, but we should be forced to design a simulation method to account for the rate of births and deaths and the effects of these two processes on the degree distribution.

The aforementioned features might lead to a situation in which the infection dynamics would modify the underlying structure of the network through which the disease is being spread. Therefore, $\lambda_c$ could also vary as one expects it to be intrinsically related to the first two moments of a seemingly time-dependent degree distribution. The reason why we consider the distribution of contacts per unit time as heterogeneous, even for the current airborne-transmitted disease is based on the observation that the number of contacts a person can have per unit of time is subjected to two sources of heterogeneity. Firstly, what we can call \emph{geo-demographic, macroscopic}  heterogeneity, in which the number of contacts depends on the population density in the region in which an individual inhabits. Secondly, at a more \emph{individual, microscopic} level, the heterogeneity arises because the number of contacts depends, in a region of constant population density (i.e., a town or neighborhood in a city), on the daily activity pattern of the individual within that region. These two factors define, ultimately, the function $P(k)$. Having that said, the assumption implicitly incorporated in the first equation of system\ (\ref{h}) does not hold. Note that this equation implies that the connectivity of individuals is hereditary and therefore that the number of births within each $k$ class equals the birth rate times the number of individuals within each $k$ class, $N_k=P(k,t)N$.

The above situation would be equivalent to assume that the dynamics of the disease being studied is the only one that influences the demographic structure of a population, which is not true since it is clear that there are countless cultural, economic and social factors that ultimately define the above two levels of heterogeneity. We therefore assume in what follows that the newborns of each generation are distributed among the $k$ classes according to an invariant distribution function, which we further assume to be the initial degree distribution of the original network: $P(k,t_o)$. As we shall see, this assumption, besides being more plausible, has the advantage that makes the connectivity distribution to be roughly stable, and so will be the critical value $\lambda_c$.

So, we have the following reformulation of the system of differential equations\ (\ref{h}):
\begin{equation}
 \frac{dU_k(t)}{dt}=bP(k,t_o)N-\lambda k\Theta(t) U_k(t)-\mu U_k(t),\nonumber
 \end{equation}
\begin{equation}
 \frac{dL_k(t)}{dt}=(1-p) \lambda k\Theta(t) U_k(t)-(\mu+r) L_k(t),
\end{equation}
\begin{equation}
 \frac{dT_k(t)}{dt}=p \lambda k\Theta(t)U_k(t)+rL_k(t)-(\mu+\mu_{tb}) T_k(t),\nonumber
\end{equation}
with the definition of the number of individuals in each class of connectivity:
\begin{equation}
N_k(t)=N(t)P(k,t),
\end{equation}
and inside each class:
\begin{equation}
U_k(t)=N_k(t)u_k(t),\nonumber
\end{equation}
\begin{equation}
L_k(t)=N_k(t)l_k(t),
\end{equation}
\begin{equation}
T_k(t)=N_k(t)t_k(t).\nonumber
\end{equation}
Now the total population within each connectivity class verifies:
\begin{equation}
 \frac{dN_k(t)}{dt}=bNP(k,t_o)-\mu NP(k,t) -\mu_{tb}T_k(t),
\end{equation}
so that, if we add in $k$, the last modification has no effect on the variation of the total volume of the population. The temporal evolution of the degree distribution is now given as: 
\begin{equation}
 \frac{dP(k,t)}{dt}=b\left[P(k,t_o)-P(k,t)\right]-P(k,t)\mu_{tb}\left[t_k(t)-\langle t_k\rangle (t)\right].
 \label{i}
\end{equation} 
Finally, writing the equations in terms of the densities we get
\begin{equation}
 \frac{du_k(t)}{dt}=b\frac{P(k,t_o)}{P(k,t)}\left(1-u_k(t)\right)-u_k(t)\left(\lambda k\Theta(t) -\mu_{tb}t_k(t)\right),\nonumber
 \end{equation}
\begin{equation}
 \frac{dl_k(t)}{dt}=(1-p) \lambda k\Theta(t) u_k(t)-\left(b\frac{P(k,t_o)}{P(k,t)}+r\right) l_k(t)+\nonumber
\end{equation}
\begin{equation}
+\mu_{tb}l_k(t)t_k(t),
 \label{j}
\end{equation}
\begin{equation}
 \frac{dt_k(t)}{dt}=p \lambda k\Theta(t)u_k(t)+rl_k(t)-\left(b\frac{P(k,t_o)}{P(k,t)}+\mu_{tb}\right)t_k(t)+\mu_{tb}t_k(t)^2.\nonumber
\end{equation}

\subsection{Characterization of the equilibrium points.}

The previous set of differential equations tells us how the different densities of interest evolves within each connectivity class. Their corresponding macroscopic quantities are defined as
\begin{eqnarray}
\langle u\rangle(t)=\sum{k}{P(k,t)u_k(t)},\nonumber\\
\langle l\rangle(t)=\sum{k}{P(k,t)l_k(t)},\label{k}\\
\langle t\rangle(t)=\sum{k}{P(k,t)t_k(t)},\nonumber
\end{eqnarray}
where $\langle u\rangle(t), \langle l\rangle(t)$ and $\langle t\rangle(t)$ are the mean densities of healthy, latent and sick individuals, respectively.

Let us now go one step further and characterize the equilibrium points. The magnitudes of interest are the average densities, so that an equilibrium point  $(\langle u\rangle^*,\langle l\rangle^*,\langle t\rangle^*)$ must verify by definition:
\begin{equation}
 \left(\frac{d\langle u\rangle^*}{dt},\frac{d\langle l\rangle^*}{dt},\frac{d\langle t\rangle^*}{dt}\right)=(0,0,0).\nonumber
\end{equation}
We also impose a further constraint which is that the degree distribution of the network is stationary, that is:
\begin{equation}
 \frac{dP(k)^*}{dt}=0\qquad\qquad\forall k.\nonumber
\end{equation}
At this point one must ask whether macroscopic stability also implies stability within each connectivity class. The answer is yes, if we also demand stability of the degree distribution. Admittedly, if we equate expression\ (\ref{i}) to zero and solve for the stationary $P(k,t)^*$ we get:
\begin{equation}
 P(k,t)^*=\frac{bP(k,t_o)}{b+\mu_{tb}(t_k^*-\langle t\rangle^*)},\nonumber
\end{equation}
which shows that this value depends on the microscopic scale $t_k$. Therefore, the stability of the degree distribution imposes a stationary condition on $t_k$ for all $k$, which in its turns extends to the other densities $l_k^*$ and $u_k^*$. Hence, we have:
\begin{equation}
 \left(\frac{du_k^*}{dt},\frac{dl_k^*}{dt},\frac{dt_k^*}{dt}\right)=(0,0,0)\quad\quad \forall k.\nonumber
\end{equation}
The above condition is trivially satisfied for the solution $(u_k^*,l_k^*,t_k^*)=(1,0,0)$ $\forall k$, which leads to a degree distribution exactly as the initial distribution. We next analyze the stability of this solution, which shall allow us to characterize the epidemic threshold.

\subsection{Epidemic threshold}

As stated before, in this section we will study the stability of the solution  $(u_k^*,l_k^*,t_k^*)=(1,0,0)$ $\forall k$. At this point, as no latent or infected individuals are introduced in the network, the degree distribution does not change in time; so that $P(k)^*=P(k,t)=P(k,t_o)$. This situation allows to work with the system of differential equations given by \ (\ref{g}) instead of working with the more general case given by the system\ (\ref{j}).

\subsubsection{Case $p=1$}

For simplicity and to gain some preliminary insight into the problem, we first study the case  $p=1$, which means that there is no latent phase (i.e., the latent subpopulation disappears, $l_k=0$ $\forall k$). Using $u_k+t_k=1$ we get,
\begin{equation}
\frac{du_k}{dt}=b-u_k(b+\lambda k \Theta -\mu_{tb})-\mu_{tb}t_k^2,\nonumber
\end{equation}
where we have omitted temporal dependences, as we will do from now on. Looking for the stationary solution, we have that the condition $\frac{du_k}{dt}=0$ implies:
\begin{equation}
u_k=-\left(\frac{1}{2\mu_{tb}}\right)\left(b+\lambda k \Theta -\mu_{tb}\pm\sqrt{(b+\lambda k \Theta -\mu_{tb})^2+4b\mu_{tb}}\right),\nonumber
\end{equation}
from which the meaningful solution is the one with the negative sign. The previous expression is consistent with the meaning of $u^*$ since we recover the expected result $u^*=1$ when $\Theta=0$. Moreover, if we calculate the derivative with respect to $\Theta$ we get:
\begin{equation}
\frac{du_k^*(\Theta)}{d\Theta}=\frac{\lambda k}{2\mu_{tb}}\left(-1+\frac{b+\lambda k \Theta -\mu_{tb}}{\sqrt{(b+\lambda k \Theta -\mu_{tb})^2+4b\mu_{tb}}}\right)<0,\nonumber
\end{equation} 
which guarantees that $u^*$ will always be less than unity and therefore is a real, valid solution. The study of the value of $\Theta$ in the steady state help us to identify the epidemic threshold. We write:
\begin{equation}
\Theta^*=\frac{1}{\langle k\rangle }\sum_kkP(k)t_k^*=1-\frac{1}{\langle k\rangle}\sum_kkP(k)u_k^*,\nonumber
\end{equation} 
which, after substituting $u^*_k$ for its value, leads to:

\begin{equation}
\Theta^*=f(\Theta)=\frac{1}{2}-\frac{b}{2\mu_tb}+\frac{\lambda\langle k^2 \rangle}{2\mu_{tb}\langle k \rangle}\Theta-\nonumber
\end{equation} 
\begin{equation}
-\frac{1}{2\mu_{tb}\langle k \rangle}\sum_kkP(k)\sqrt{(b+\lambda k \Theta -\mu_{tb})^2+4b\mu_{tb}}.\nonumber
\end{equation} 

The graphical interpretation of the above equation indicates that the existence of an equilibrium point in which $\Theta^*>0$ is equivalent to the existence of a point at which $f(\Theta)$ crosses the bisector of the first quadrant. Evaluating the second derivative of $f(\Theta)$ one gets:
\begin{equation}
\frac{d^2f(\Theta)}{d\Theta^2}=\frac{-2b\lambda^2}{\langle k \rangle}\sum_k\frac{P(k)k^3}{\left[(b+\lambda k \Theta -\mu_{tb})^2+4b\mu_{tb} \right]}<0,\nonumber
\end{equation}
which ensures that the condition for the existence of such intersection is reduced to:
\begin{equation}
\left(\frac{df(\Theta)}{d\Theta}\right)_{\Theta=0}=\frac{\lambda\langle k^2 \rangle}{\langle k \rangle}\frac{1}{b+\mu_{tb}}=1,\nonumber
\end{equation}
from which the epidemic threshold is derived as:
\begin{equation}
\lambda_c=\frac{(b+\mu_{tb})\langle k \rangle}{\langle k^2 \rangle}.\nonumber
\end{equation}
Note that apart from the factor $(b+\mu_{tb})$, the previous result, formally coincides with the epidemic threshold of the SIR model.

\subsubsection{Case $p\neq1$}

This is a somewhat more involved case. For structured populations, the resolution of the system of differential equations\ (\ref{j}) cannot be done explicitly. We next find the epidemic threshold for the case $p\neq1$ using two approaches. On one hand, we study the time derivative of $\Theta$.  On the other hand, we will also make use of the singularity of the Jacobian at the point $(u_k,l_k,t_k)=(1,0,0)$ to argue that the expression for the critical threshold is given by:
\begin{equation}
 (\lambda)_c=\frac{\langle k \rangle}{\langle k^2 \rangle}\frac{(r+b)(\mu_{tb}+b)}{pb+r}.
\end{equation}
\paragraph{Time evolution of $\Theta$ in populations with a low number of sicks.}

A first approach to characterize the epidemic threshold in heterogeneous networks when $p\neq1$ is to study the sign of the derivative of $\Theta$ at the onset of an epidemic outbreak. We consider an initially healthy population in which a small proportion of infectious individuals is introduced so that $t_k<<1$ $\forall k$. The derivative of $\Theta$ is:
\begin{equation}
 \left(\frac{d\Theta}{dt}\right)_{\Theta\sim0}=\frac{\sum_k{P(k)k\frac{dt_k}{dt}}}{\langle k \rangle}+\frac{\sum_k{t_kk\frac{dP(k)}{dt}}}{\langle k \rangle}-\nonumber
\end{equation}
\begin{equation}
-\left[\frac{\sum_k{P(k)kt_k}}{\langle k \rangle}\right]\left[\frac{\sum_k{k\frac{dP(k)}{dt}}}{\langle k \rangle}\right]\nonumber
\end{equation}
which, after substitution of the values of the derivatives of $P(k,t)$ and $t_k(t)$ leads to:
%
\begin{equation}
 \left(\frac{d\Theta}{dt}\right)_{\Theta\sim0}=\frac{\sum_k{P(k)kl_k}}{\langle k \rangle}+p\lambda\Theta\frac{\sum_k{P(k)k^2u_k}}{\langle k \rangle}-\left(b+\mu_{tb}\right)\Theta+\mu_{tb}\Theta^2.\nonumber
\end{equation}
%
At this point we make two simplifications. The first and most easily justifiable is to neglect the term $\Theta^2$. The second is related to the presence of $l_k$ in the above equation, that we have to transform in a dependency with respect to $t_k$. Specifically, we assume to be sufficiently close to the stationary point $(u_k,l_k,t_k)=(1,0,0)$ as to be able to assume that the three derivatives vanish. In other words, and focusing our attention on latent and sick classes, we assume that:
\begin{equation}
 \left(\frac{dl_k}{dt}\right)_{\Theta\sim0}=(1-p) \lambda k\Theta u_k-(b+r) l_k+\mu_{tb}l_kt_k\simeq0,\nonumber
\end{equation}
\begin{equation}
\left(\frac{dt_k}{dt}\right)_{\Theta\sim0}=p \lambda k\Theta u_k+rl_k-(b+\mu_{tb})t_k+\mu_{tb}t_k^2\simeq0,\nonumber
\end{equation}
from which: 
\begin{equation}
l_k=\frac{\left(1-p)(b+\mu_{tb}\right)t_k-(1-p)\mu_{tb}t_k^2}{r+pb-\mu_{tb}pt_k}=\frac{\left(1-p)(b+\mu_{tb}\right)}{r+pb}t_k + O(t_k^2),\nonumber
\end{equation}
which allows to express the derivative of $\Theta$ as:
\begin{equation}
 \left(\frac{d\Theta}{dt}\right)_{\Theta\sim0}=\Theta\left[\frac{r\left(1-p)(b+\mu_{tb}\right)}{r+pb}-(b+\mu_{tb})+\right .\nonumber
\end{equation}
\begin{equation}
 \left .+p\lambda\frac{\sum_k{P(k)k^2u_k}}{\langle k \rangle}\right].\nonumber
\end{equation}
In the limit $u_k\simeq1$ $\forall k$ the third term within brackets is the ratio $\langle k^2 \rangle/\langle k \rangle$, from which the epidemic threshold condition may be derived as:
\begin{equation}
 \left(\frac{d\Theta}{dt}\right)_{\Theta\sim0}=\Theta\left[\frac{r\left(1-p)(b+\mu_{tb}\right)}{r+pb}-(b+\mu_{tb})+p\lambda_{c}\frac{\langle k^2 \rangle}{\langle k \rangle}\right]=0,\nonumber
\end{equation}
finally leading to the expected expression for the threshold:
\begin{equation}
\lambda_c=\frac{\langle k \rangle}{\langle k^2 \rangle}\frac{(r+b)(\mu_{tb}+b)}{pb+r}.
\label{threshold}
\end{equation}
\paragraph{Analysis of the Jacobian}

While for well-mixed populations the condition of singularity of the Jacobian allows to get the epidemic threshold in a straightforward way, for heterogeneous populations the analysis of the Jacobian is a difficult task because $\Theta$ is a function of each and every one of the $t_k$'s. This translates into the need of calculating a determinant whose order is three times the number of connectivity classes. What we can reasonably do is to verify is the threshold condition is verified for systems in which there are two or three different connectivity classes.
In the first case in which only two different classes of connectivity exist, the Jacobian is just a quite distasteful 6x6 determinant that, after some cumbersome and lengthy algebra, can be reduced to the expression:
\begin{equation}
J=b^2(b+r)(b+\mu_{tb})\left[(b+r)(b+\mu_{tb})+\frac{\lambda\langle k^2\rangle}{\langle k\rangle}(pb+r)\right],\nonumber
\end{equation}
which equated to zero leads again to the previously obtained expression for the epidemic threshold.
If we instead consider a population with three degree classes, the algebraic complexity of the problem largely increases as we now have to solve a determinant of size 9x9. However, we can proceed as before getting the following expression for the Jacobian:
\begin{equation}
J=b^3(b+r)^2(b+\mu_{tb})^2\left[(b+r)(b+\mu_{tb})+\frac{\lambda\langle k^2\rangle}{\langle k\rangle}(pb+r)\right].\nonumber
\end{equation}
This leads us to the sensible conjecture that increasing the number of connectivity classes does not add new roots to the Jacobian, but it only would increase the degeneracy of the non-interesting solutions $b=0$, $b=-r$ and $b=-\mu_{tb}$.

\subsection{Numerical simulations}

When designing numerical simulations to inspect the dynamics of the system under study, we have two difficulties not previously addressed in the literature. These numerical issues with which we have to deal come from the fact that we have a system that is simultaneously open and structured. As a result of dealing with an open system, new individuals are being added to the population at a rate given by the birth rate. Additionally, these new individuals must enter the network of contacts with a predefined connectivity. While deciding how many nodes our new individuals connect to is not a problem, it certainly is to decide what are those nodes the newcomers will be linked to, as this will impact the degree distribution in a nontrivial way. This is an unavoidable numerical complication that we should face relentlessly if the analytical calculations are to be compared with Monte Carlo simulations.

\begin{figure}
\epsfig{file=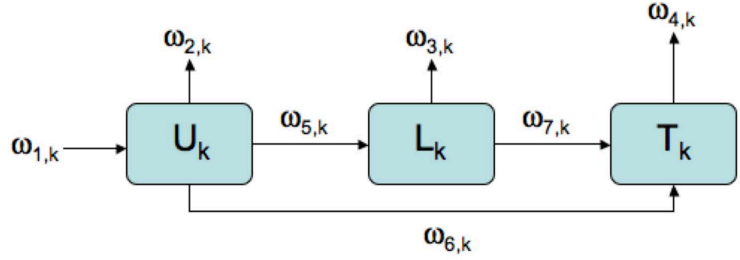,width=4.1in,angle=0,clip=1}
\caption{Set of allowed transitions in the epidemic model within each connectivity class.}
\label{figure1}
\end{figure}

To this end, we have adapted a simulation method based on transition probabilities first proposed in \cite{PRE} for SIR models in complex networks. The numerical approach considers all transitions between states that take place during the dynamical evolution of the subpopulations, defined by the system of differential equations\ (\ref{j}). When dealing with structured populations, these transition rates depend, in general, on the connectivity class within which they occur. Moreover, within each $k$-class seven transitions are possible (see Fig.\ \ref{figure1}):
\begin{description}
 \item [-] Birth of healthy individuals,
 \item [-] Natural death of healthy individuals,
 \item [-] Natural death of latently infected individuals,
 \item [-] Natural or disease-related death of sick individuals,
 \item [-] Transition from a healthy to the latent state,
 \item [-] Transition from a healthy to the sick (infectious) state,
 \item [-] Transition from a latent to the sick (infectious) state.
 \end{description}
Each of these transitions is characterized by a characteristic transition rate $\omega_{i,k}$ that can be directly derived from the system of equations that characterizes the rate at which they occur within the class $k$ as:

\begin{center}
\begin{tabular}{c}
$\omega_{1,k}=bNP(k,t_o)$\\
$\omega_{2,k}=\mu N P(k,t)u_k$\\
$\omega_{3,k}=\mu NP(k,t)l_k$\\
$\omega_{4,k}=(\mu+\mu_{tb})NP(k,t)t_k$\\
$\omega_{5,k}=(1-p)\lambda kNP(k,t)u_k\Theta$\\
$\omega_{6,k}=p\lambda kNP(k,t)u_k\Theta$\\
$\omega_{7,k}=r NP(k,t)l_k$\\
\end{tabular}
\end{center}
Similarly, we define the sum of all these transition rates as the average rate at which \emph{one} transition (of any kind) occurs:
\begin{equation}
\Omega=\sum_{i,k}{\omega_{i,k}}.
\end{equation}
This average transition rate in its turn defines the characteristic or average time $\tau$ elapsed between any two consecutive transitions, the latter being defined as the inverse of $\Omega$:
\begin{equation}
\tau=\frac{1}{\Omega}.
\end{equation}
Given the previous definitions, the Monte Carlo algorithm is implemented in such a way that at each MC step (of duration $\tau$) one single transition takes place. Finally, the probability $\Pi_{i,k}$ that a given transition actually happens, is calculated as:
\begin{equation}
\Pi_{i,k}=\frac{\omega_{i,k}}{\Omega}=\tau\omega_{i,k},
\end{equation}
that determines which of all possible transitions is realized at each time step $\tau$.

\begin{figure}
\epsfig{file=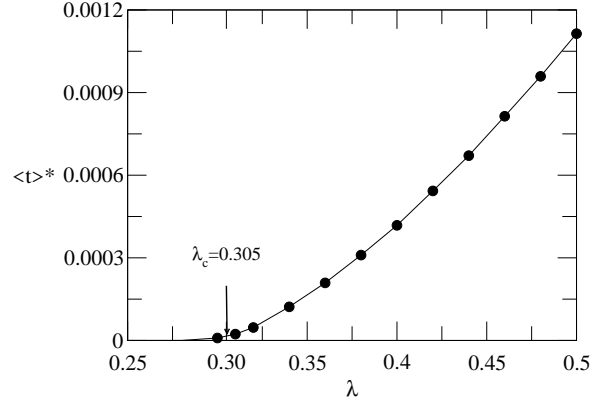,width=3.5in,angle=0,clip=1}
\caption{Stationary proportion of sick individuals as a function of $\lambda$ ($\in[0.25,0.5]$) for $p=0.07$, $r=0.002$, $\mu_{tb}=0.2$, $\mu=0.009$ and $b=0.01$. The arrow marks the position of the epidemic threshold ($\lambda_{c}=0.305$) as given by analytical calculations using the previous values for the parameters and the degree distribution. Error bars are smaller than symbol size. The initial size of the network, characterized by a degree distribution $P(k,t_o)\sim k^{-3}$, is $N_0=10^6$. Each point is an average over at least 1000 values of $\langle t\rangle^*$.}
\label{figure2}
\end{figure}

We have made extensive numerical simulations of the model starting from an initial population made up of $N\simeq10^6$ individuals, whose network of contacts follows an initial degree distribution $P(k)\simeq k^{-3}$. Moreover, every newborn joins the system with a degree that verifies the same connectivity distribution. As for the values of the parameters of the dynamics, and thinking of typical values for persistent diseases, we have set the following values: $\mu=0.009 \quad \text{years}^{{-1}}$, $b=0.010 \quad \text{years}^{{-1}}$, $\mu_{Tb}=0.200 \quad \text{years}^{{-1}}$, $r=0.002 \quad \text{years}^{{-1}}$, and $p=0.070$. Demographical parameters $b$ and $\mu$ are roughly those of a country like Spain, while the parameters $p$ and $r$ are in the range of typical values for the case of tuberculosis. $\mu_{tb}$ has been chosen attending to numerical convenience (tuberculosis reaches a disease-related mortality rate that can be as large as 0.8). On the other hand, we note that a numerical criterion to define stationarity should also be adopted. In our simulations, we first let the system evolve for $4500$ years and later take averages in a window of $10^6$ Monte Carlo steps (which corresponds, roughly, with a temporal lapse of 100 years), for the mean densities of healthy, latent and sick individuals defined in\ (\ref{k}). This is a long enough time and ensures that all of the outputs to the left of the threshold are stabilized to the state $(1,0,0)$ (this is achieved almost surely already for $t\leq4000$ years).

With the values for the parameters as specified above, the epidemic threshold is $\lambda_c=0.305$. In Fig.\ \ref{figure2}), we have plotted the stationary proportion of sick individuals for values of the probability of transmission $\lambda$ in the interval $[0,0.5]$. As can be seen from the figure, the numerical and analytical results are in a reasonable agreement, despite the numerical challenges of simulating an open system in which the dynamics evolves on top of a complex topology. This indicates that the numerical method is accurate enough as to be used in situations where analytical predictions are not at hand. 

\begin{table}
\caption{Dependency of the epidemic thresholds on the initial size of the network. Analytical values are obtained from Eq.\ (\ref{threshold}) using the moments of the distribution generated numerically.}
\begin{tabular}{ccc}\hline
\quad $N_{o}$\quad &\quad Analytical\quad &\quad Numerical \quad \\
 &$\lambda_{cA}$ &$\lambda_{cN}$\\\hline
1000&	0.52903&	0.46968 \\\hline
10000&	0.42559&	0.37595 \\\hline
50000&	0.37620&	0.33088 \\\hline
100000&	0.35854&	0.31926 \\\hline
\end{tabular}
\label{default}
\end{table}

However, it is possible to carry out numerical simulations using a variation of the algorithm above in order to improve the accuracy in the determination of the epidemic threshold $\lambda_{c}$. In this variation, instead of using a criterion for stationarity, we focus our attention in the vicinity of the critical value. More specifically, we first evaluate analytically the value for $\lambda_{c}$ and start the simulation there (obviously, if we don't have an analytical hint, the simulation can be started at any value of $\lambda$). At each realization, we expect 6000 years for an eventual arrival of the system to the state $(\langle u \rangle,\langle l \rangle, \langle t \rangle)=(1,0,0)$. In the case that this state is not reached in that time, we assume that we are to the right of the critical point and so, we move to the left in $\lambda$ just a little quantity $\delta\lambda=0.01$. If, on the contrary, a minimum number of realizations (we used 10 in our simulations) stabilize at state $(\langle u \rangle,\langle l \rangle, \langle t \rangle)=(1,0,0)$, we assume to be to the left of the critical point and, consequently, we perform a $\delta\lambda$ switch to the right. Each time that such kind of flip-flop algorithm (a sort of bisection method) changes direction, we divide by two the value of $\delta\lambda$ until the desired precision in $\lambda_{c}$ is obtained. Using this numerical approach, we have numerically calculated the values of $\lambda_c$ for different system sizes. The results are reported in Table\ \ref{default}. As expected from finite-size effects, the larger the size of the population, the smaller the absolute error between numerical and analytical thresholds is. Moreover, the larger the system size the smaller the epidemic threshold, which eventually should vanish in the thermodynamic limit.

\section{Conclusions.}

We have discussed a model for the spreading of persistent infections in complex heterogeneous populations. The framework extends the epidemiological picture proposed in previous works.  Our approach is particularly suited for diseases like Tuberculosis, which shows large latency periods. The latent period results from the dynamical equilibrium that is established between the bacterium and the host's immune system, so that the host might not become infectious during its lifetime. These particular features makes it compulsory to work with an open system where newborns are continuously introduced in the population and individuals might die due to causes different from the disease itself. By assembling a model with all these ingredients, we have shown analytically that the epidemic threshold is proportional to the ratio between the first and second moments of the degree distribution. Therefore, our results point in the same direction that those obtained for the SIS and SIR model on top of the same topologies -the virtually unbounded connectivity fluctuations play a key role in the infection dynamics enhancing the epidemic incidence and lowering the epidemic threshold.

From another point of view, we have developed a method well suited to numerically explore the dynamics of the system under study. In particular, we have been able to deal with the new challenge of having a system where the number of individuals in the population is not constant and, moreover, are connected following a given degree distribution (the well-mixed case is not challenging). Although we have applied the numerical approach to our particular system, it is worth stressing that it is general and can be applied to any problem for which transition rates between different classes and states are known. The results obtained agree well with the analytical estimates with the additional advantage that the whole phase diagram can be explored. 

Finally, it is also worth mentioning that the model discussed here is probably the simplest one may devise for the spreading of persistent infections in structured populations. However, despite the recent progresses in modeling disease contagion dynamics and pandemic outbreaks, the kind of spreading phenomena analyzed here is one of the issues that have remained less explored. Our aim is to take a first step towards more realistic modeling of persistent diseases. We have left for future investigation possible extensions of the current model that take into account the influence on the dynamics of vaccination, prophylaxis and recovery rates, as well as the effects of genetic heterogeneity in the pathogen but also in the host \cite{MB}. 

\begin{acknowledgments}

We acknowledge financial support of the Government of Arag\'on through Grant PI038/08. This work has also been partially supported by MICINN through Grants FIS2008-01240 and FIS2009-13364-C02-01. 

\end{acknowledgments}

\end{document}